\documentclass[12pt]{article}
\usepackage{amsmath}
\usepackage{amsfonts}
\usepackage{amssymb}
\usepackage{amstext}
\usepackage{graphicx}
\usepackage{epsfig}%
\title{\centerline
\bf Possible potentials responsible for stable circular relativistic orbits}
\bigskip
\author{Prashant Kumar$^\dagger$, Kaushik Bhattacharya$^\ddagger$
\thanks{email:\,\,$^\dagger$kprash@iitk.ac.in, $^\ddagger$kaushikb@iitk.ac.in}
\\
\normalsize
Department of Physics, Indian Institute of Technology, Kanpur\\ 
\normalsize
Kanpur 208016, India
}
\begin{document}
\maketitle
\begin{abstract}
Bertrand's theorem in classical mechanics of the central force fields
attracts us because of its predictive power. It categorically proves
that there can only be two types of forces which can produce stable,
circular orbits. In the present article an attempt has been made to
generalize Bertrand's theorem to the central force problem of
relativistic systems. The stability criterion for potentials which can
produce stable, circular orbits in the relativistic central force
problem has been deduced and a general solution of it is presented in
the article. It is seen that the inverse square law passes the
relativistic test but the kind of force required for simple harmonic
motion does not. Special relativistic effects do not allow stable,
circular orbits in presence of a force which is proportional to the
negative of the displacement of the particle from the potential
center.
\end{abstract}
\section{Introduction}
The central force problem in
non-relativistic classical mechanics is one of the most useful topics
in physics. Closely linked with the central force problem is the
Keplerian orbit theory which is a cornerstone for understanding
planetary motions in the solar system or motion of electrons near the
nucleus. In classical mechanics there is an important theorem called
the Bertrand's theorem which proposes that there can only be two types
of central potentials, the Coulomb type and the simple harmonic type,
which can produce stable, circular orbits for particles moving around
the potential source. A good presentation of the Bertrand's theorem can be 
found in Ref.~\cite{goldstein}.  The present article tries to
generalize the results of Bertrand's theorem when the orbiting
particle can have relativistic velocities.
 
In this article we first set up the relativistic orbit equation for a
particle in a central potential presumed to be dependent on the radial
coordinate only. The relativistic central force orbits were previously
studied in Refs.~\cite{boyer, tork, reut, frommert}.  A brief
description of the central force problem in a relativistic setting in
a Coulomb potential was presented in the book on classical theory of
fields by Landau and Lifshitz \cite{landau}. Before one starts the
main analysis about the stability of orbits of relativistic particles
in a central force potential it is better to specify the assumptions
one makes in arriving at definite results. In the present article we
use the same assumptions and the approximations as utilized by Boyer
in Ref.~\cite{boyer} and Landau in Ref.~\cite{landau}.  In the
specific references cited above, none of them present a Lorentz
covariant treatment of the relativistic central force problem. The
main reason being that all of them assumes a central potential $V(r)$
where $r=|{\bf r}|$ is the distance between the source and the
orbiting particle. The form of the potential only depends on the
position coordinates of the orbiting particle. The form of $V(r)$ is
not Lorentz covariant. In such cases the results of the whole analysis
is valid in a particular frame where the origin of the coordinate
system coincides with the potential center.

The references cited above assumes the particle which produces the
potential $V(r)$ to be static in the specific coordinate system
utilized by the observer. If the source of the potential does not have
any velocity then the retarded nature of the interactions, owing to
the finite velocity of light, does not complicate the calculation of
the orbit of the relativistic particle. A specific example will make
the point clear.  In classical electrodynamics if the source of the
Coulomb potential $V(r)$ moves with a velocity ${\bf v}_s$ and the
orbiting particle has a velocity ${\bf v}$ then the potential $V(r)$
gets a relativistic correction. The magnitude of the lowest order
relativistic correction to the Coulomb potential was calculated by
Darwin in 1920 and it looks like
$$\frac{V(r)}{2c^2}\left[{\bf v}_s \cdot {\bf v} + \frac{({\bf
      v}_s\cdot{\bf r})({\bf v}\cdot{\bf r})}{r^2}\right]\,.$$ For a
better understanding of the Darwin correction one can look at
Ref.~\cite{jackson}. In our case ${\bf v}_s=0$ and consequently there
will be no relativistic modification of $V(r)$. More over we do
not consider any general relativistic effects due to $V(r)$ into
account. We briefly comment on the general relativistic generalization
of the central force problem in section \ref{reltd}.  In the present
article the background space-time is assumed to be flat. 

In the article it will be shown that the stability condition of the
perturbed orbits around a stable circular orbit gives rise to a
non-linear differential equation for the central potential. The
Newtonian or the Coulomb potential satisfies the resulting
differential equation with some restrictions on the possible value of
the angular momentum of the orbiting particle. Except the Newtonian
potential solution we present a more general solution of the
differential equation for the potential which can give rise to stable,
circular orbit for relativistic particles. This solution gives rise to
a force which is not common in physics except its Newtonian inverse 
square law limit. The equation of the orbit of a relativistic particle
in such a non-trivial force shows that the orbit will precess and the
precession angle can be calculated.

Unlike the non-relativistic case, in the relativistic case there exist
no radial effective potential minimizing which we can obtain the
radius of a circular orbit. In the relativistic case a first order
perturbation from a circular orbit is enough to determine the
stability criterion of the orbit. In the non-relativistic case one
uses higher order perturbations from a circular orbit to specify the
form of the potential. In the relativistic case the general solution
of the form of the potential from first order perturbation from
circular orbit is such that all higher order corrections becomes
irrelevant. As a consequence of this fact the general form of the
potential which can produce stable, circular orbits for relativistic
particles contains more parameters than the corresponding expressions
of non-relativistic potentials.

The material in the article is presented in the following manner. The
second section sets the conventions and derives the orbit equation of
a relativistic particle in a central orbit. Section \ref{stability}
generalizes the Bertrand's theorem for the relativistic case. In this
section the stability condition for the circular orbits will be
interpreted as a non-linear differential equation for the
potential. The solutions of the non-linear stability equation will
also be derived in section \ref{stability}. In section \ref{reltd} the
connection of the present work with some related works which were
existing in the literature are discussed. This section gives a wider
view for the readers who really want to understand the stability of
orbits in special relativity and general relativity. The last section
\ref{concl} summarizes the important points presented in the article.
\section{The orbit equation}
\label{orbit}
In this section we derive the orbit equation of the relativistic
particle in presence of a potential $V(r)$ which is purely a function
of the radial coordinate.  The Lagrangian of a relativistic particle
of mass $m$ in presence of a continuous radial potential $V(r)$ is
\begin{eqnarray}
{\mathcal L} &=& -mc^2\sqrt{1-{v^2}/{c^2}} - V(r)\,,
\label{lagrange} 
\end{eqnarray}
where the velocity of the particle ${\bf v}$ in plane polar
coordinates is given as
$${\bf v}=\dot{r}\hat{e}_r + r\dot{\theta}\hat{e}_\theta\,,$$ where
$\hat{e}_r$ , $\hat{e}_\theta$ are the mutually orthogonal unit
vectors along the radial and the angular directions.  The form of the
Lagrangian in Eq.~(\ref{lagrange}) immediately shows that the angular
momentum
\begin{eqnarray}
L=\frac{\partial {\mathcal L}}{\partial \dot{\theta}}=
mr^2 \gamma \dot{\theta}\,,
\label{angmom}
\end{eqnarray}
is a constant, where
$$\gamma=\frac{1}{\sqrt{1-{v^2}/{c^2}}}\,.$$ 
The total energy $E$ of the particle in presence of the potential $V(r)$ is
\begin{eqnarray}
E= m c^2 \gamma + V(r)\,.
\label{e1}
\end{eqnarray}
Although from the definition of $\gamma$ it looks like that it is a
function of $r$, $\dot{r}$ and $\dot{\theta}$ but it can be shown that
in a central force field $\gamma$ is only a function of the radial
coordinate $r$. The reason for such behavior of $\gamma$ can be
understood from the following reason. As energy and angular momentum
are constant functions of $r$, $\dot{r}$ and $\dot{\theta}$ we can use
the conservation conditions of $E$ and $L$ to re-express $\dot{r}$ and
$\dot{\theta}$ as functions of $r$, $E$ and $L$. As $E$ and $L$ are
constants so in a central force field $\dot{r}$ and $\dot{\theta}$ are
functions of $r$ alone. Consequently $\gamma$ is only a function
of $r$. In special relativity the energy of the particle in a central
potential can also be written as
\begin{eqnarray}
E=\sqrt{p^2 c^2 + m^2 c^4}+V(r)\,,
\label{energy}
\end{eqnarray}
where $p=|{\bf p}|$, 
\begin{eqnarray}
{\bf p}=m\gamma {\bf v}&=&m\gamma(\dot{r}\hat{e}_r  + 
r\dot{\theta}\hat{e}_\theta)\nonumber\\
&=&p_r \hat{e}_r + p_\theta \hat{e}_\theta\,,
\label{momentum}
\end{eqnarray}
and 
$$p_r=m\gamma \dot{r}\,,\,\,\,\,\,\, p_\theta=m\gamma
r\dot{\theta}=\frac{L}{r}\,.$$ As because
$({p_r}/{p_\theta})=({\dot{r}}/{r\dot{\theta}})$, we
have $$p_r=\frac{L}{r^2}\frac{dr}{d\theta}\,.$$ With the above
information on the various momentum components we can now rewrite 
Eq.~(\ref{energy}) as
\begin{eqnarray}
(E-V)^2 = \left(\frac{L}{r^2}\frac{dr}{d\theta}\right)^2 c^2 
+ \frac{L^2c^2}{r^2} + m^2c^4\,. 
\label{energy1}
\end{eqnarray}
Instead of $r$ we use the variable $u=\frac{1}{r}$ in terms of which
Eq.~(\ref{energy1}) becomes
$$(E-V)^2=L^2c^2\left(\frac{du}{d\theta}\right)^2 + u^2L^2c^2 + m^2c^4\,.$$ If
we differentiate the last equation with respect to $\theta$ and then
divide the resulting equation by $du/d\theta$ we obtain the desired
equation of the orbit of a particle of mass $m$ possessing momentum
${\bf p}$ moving in the presence of a general central potential $V(r)$
as
\begin{eqnarray}
\frac{d^2 u}{d \theta^2} + u = \frac{(V-E)}{L^2 c^2}\frac{dV}{du}\,.
\label{orbit_eqn}
\end{eqnarray}
Using Eq.~(\ref{e1}) we can rewrite the above equation in the form
\begin{eqnarray}
\frac{d^2 u}{d \theta^2} + u =
-\frac{m\gamma}{L^2}\frac{dV}{du}\,.
\label{gorbit_eqn}
\end{eqnarray}
 Writing $L=\gamma\ell$, where
$\ell=mr^2\dot{\theta}$ is the non-relativistic angular momentum, the
above equation in the non-relativistic limit ($\gamma \to 1$)
transforms exactly to the form we get in a conventional
non-relativistic treatment of the problem as given in Ref.~\cite{goldstein}.
\section{Circular, Stable closed orbits}
\label{stability}
Lets define 
\begin{eqnarray}
J(u)\equiv \frac{(V-E)}{L^2c^2}\frac{dV}{du}\,.
\label{defj}
\end{eqnarray}
Suppose Eq.~(\ref{orbit_eqn}) admits a circular orbit of radius
$r_0=1/u_0$.  For small perturbations around this circular orbit we
can Taylor expand $J(u)$ around $u_0$. Keeping up to first order terms
in the perturbation of $u$ we get
\begin{eqnarray}
J(u)=J(u_0) + (u-u_0)\left(\frac{dJ}{du}\right)_{u_0}\,.
\label{pertbj}
\end{eqnarray}
Noting that $J(u_0)=u_0$ for the circular orbit, we can now write 
Eq.~(\ref{orbit_eqn}) as
\begin{eqnarray}
\frac{d^2 u}{d \theta^2} + (u-u_0) = (u-u_0)\left(\frac{dJ}{du}\right)_{u_0}\,.
\nonumber
\end{eqnarray}
If we define $x \equiv u-u_0$ then the above equation can be written as
\begin{eqnarray}
\frac{d^2 x}{d \theta^2} + \zeta^2 x = 0\,,
\label{stbld}
\end{eqnarray}
where $\zeta^2$ is defined as 
\begin{eqnarray}  
\zeta^2 \equiv 1- \left(\frac{dJ}{du}\right)_{u_0}\,.
\label{betasqrd}
\end{eqnarray}
From Eq.~(\ref{stbld}) it is clear that if the orbit of the
relativistic particle in a general central potential has to be stable
then $\zeta^2 > 0$ and if the orbit has to be closed then $\zeta$ must
be a rational number.
\subsection{A differential equation for the potential $V(r)$ producing 
stable and closed circular orbits}
The rational number $\zeta$ as predicted, in Eq.~(\ref{betasqrd}),
from the stability criterion of closed circular orbits in the central
force problem is an interesting input in the theory.  The interesting
property about this rational number is that it is a constant and so it
does not depend on the details of the orbit which one tries to
perturb. The reason for the constancy of $\zeta$ is the following. For
any circular orbit with radius $r_0$ a specific $\zeta$ specifies the
number of undulations of the perturbed orbit. If $\zeta$ is a rational
number then the number of undulations of the perturbed orbit will be
such that they form a closed geometrical structure. Now suppose one
takes another circular orbit of radius $r_0+\delta r$ where $\delta r
\ll r_0$. If $\zeta$ has a different value on this orbit then the
number of undulations due to a perturbation will be different. In the
limit $\delta r \to 0$ in a continuous manner the two unperturbed
circular orbits tends to each other but the number of undulations on
the circular orbits will not match as $\zeta$ is not a continuous
variable but can only have discrete rational values. Consequently the
number of cycles of the perturbations will change discontinuously with
radius and the perturbed orbits cannot be closed at this
discontinuity. As we are only interested in stable, closed orbits we
can conclude that $\zeta$ must be a constant and not change discretely
with $r$.  The discussion on the constancy of $\zeta$ as given above
closely follows the analysis given in Ref.~\cite{goldstein} where the
author gives a nice discussion on the role of $\zeta$ in the case of
non-relativistic orbits.

As $\zeta$ is a constant and must not depend upon the choice of $u_0$
or $x$ one can interpret Eq.~(\ref{betasqrd}) as an independent
differential equation by itself,
\begin{eqnarray}  
1- \left(\frac{dJ}{du}\right)=\zeta^2\,.
\label{nbetasqrd}
\end{eqnarray}
whose solutions would give us information about the general form of the
central potential $V(r)$. Using Eq.~(\ref{defj}) we can write the last
equation as
\begin{eqnarray}
(V-E)\frac{d^2 V}{d u^2} + \left(\frac{dV}{du}\right)^2
=L^2 c^2 (1-\zeta^2)\,,
\label{maineqn}
\end{eqnarray}
which is a non-linear second order differential equation.  The right
hand side of the above equation is a constant which can be written as
\begin{eqnarray}
d=L^2c^2(1-\zeta^2)\,.
\label{ddef}
\end{eqnarray}
Eq.~(\ref{maineqn}) admits multiple solutions for $V$. The constant
$E$ is the total energy of the particle.
    
It is interesting to note that the differential equation for the
potential stemming from the stability of closed, circular orbits in
the relativistic case does not have a non-relativistic
analogue. Although the orbit equation Eq.~(\ref{gorbit_eqn}) has a
proper non-relativistic limit the same cannot be said about
Eq.~(\ref{maineqn}). The reason for such behavior can be seen clearly
if we rewrite Eq.~(\ref{maineqn}) in a slightly different way. From the 
expression of the energy of the particle in the central force field $\gamma$
can always be written as $(E-V)/mc^2$. As the total energy is a constant in the 
present case we must have $d\gamma/du=-(1/mc^2)dV/du$. Consequently
Eq.~(\ref{maineqn}) can also be written as
\begin{eqnarray}
\gamma\frac{d^2 \gamma}{du^2}+\left(\frac{d\gamma}{du}\right)^2
=\frac{L^2(1-\zeta^2)}{m^2c^2}\,,
\label{geqna}
\end{eqnarray}
which gives a differential equation of $\gamma$. The equation above
obviously does not have a well defined non-relativistic limit. The
relativistic stability condition produces an ill-defined
non-relativistic limit due to the fact that in the relativistic case
$J(u)$ as given in Eq.~(\ref{defj}) depends upon the velocity of the
orbiting particle\footnote{The $V-E$ in $J(u)$ is proportional to
  $\gamma$ which depended upon the velocity of the particle.}. In the
non-relativistic $J(u)$ was purely a function of the radial coordinate
of the orbiting particle. A perturbation from the circular orbit in
the relativistic case consists of two kinds of perturbations. One is
related to the change in position of the particle from its previous
orbit and the other is the change in velocity from the 
velocity it had previously on the circular orbit. In the
non-relativistic case only a radial perturbation from the circular
orbit fixes the shape of the stability condition. As because the
stability condition of the orbit depends upon velocity of the
relativistic particle and the corresponding non-relativistic stability
condition does not depend upon the velocity of the particle, the
non-relativistic limit of Eq.~(\ref{maineqn}) or Eq.~(\ref{geqna}) is
not well defined.

In the case of non-relativistic motion we know that the inverse square
law potential and the simple harmonic potential has the capability to
produce stable, closed circular orbits. In the present case to get the
forms of the potentials which can produce stable, closed orbits we
have to solve Eq.~(\ref{maineqn}). As it is a non-trivial equation we
will first try to see whether the the potentials which produced
stable, closed orbits in the non-relativistic regime still satisfy
Eq.~(\ref{maineqn}). Let us try to see whether any power law solution
of the form
\begin{eqnarray}
V(u)=-\alpha u^\tau\,,
\label{plaw}
\end{eqnarray}
where $\alpha>0$ satisfies Eq.~(\ref{maineqn}). In the above equation
$\alpha$ and $\tau$ are constants. If we substitute the above form of
the potential in Eq.~(\ref{maineqn}) we get
$$u^{2(\tau-1)}\left\{\alpha^2\tau(\tau-1) +
\alpha^2\tau^2\right\} -
u^{\tau-2}\left\{E\alpha\tau(\tau-1)\right\}=d\,.$$ This
directly shows that the above relation can be valid for any $u$
only if $\tau=1$, when $\alpha^2=d$ or 
\begin{eqnarray}  
L=\frac{\alpha}{c\sqrt{1-\zeta^2}}\,,
\label{lcond}
\end{eqnarray}
on using Eq.~(\ref{ddef}).  In this case we see that choosing $\tau=1$ in 
Eq.~(\ref{plaw})  we get the Coulomb or
Newtonian potential. The last equation shows that for stable,
circular orbits the particles angular momentum must satisfy some
condition. Eq.~(\ref{lcond}) implies that $\zeta^2 < 1$, and as
$\zeta^2 > 0$ for a stable orbit, we have
\begin{eqnarray}
0 < \zeta^2 < 1\,.
\label{betalim}
\end{eqnarray}
The above equation gives
\begin{eqnarray}
L > \frac{\alpha}{c}\,,
\label{ajp}
\end{eqnarray}
giving a lower bound on the angular momentum of the orbiting
particle. This lower bound of the orbiting particle was previously
obtained in a different way by T. H. Boyer in Ref.~\cite{boyer}. It
must be noted here that except $\tau=1$ no other values of $\tau$ are
allowed in the potential which can produce stable circular orbits of
relativistic particles . In non-relativistic mechanics we do also have
the harmonic-oscillator potential corresponding to $\tau=-2$ and
$\alpha<0$ in Eq.~(\ref{plaw}), but interestingly relativistic effects
forbid this value of $\tau$.
\subsection{The general solution of the differential equation for the 
potential and the nature of orbits}
We can find out the general form of the force which can produce
stable, circular relativistic orbits.  Noticing that the left hand side
of Eq.~(\ref{maineqn}) can also be written as
$$\frac{d^2}{du^2}\left[\frac{(V-E)^2}{2}\right]\,,$$
it can be easily shown that
\begin{eqnarray}
V(r)-E= - \sqrt{d\left(b+\frac{1}{r}\right)^2 + a}\,,
\label{nvform}
\end{eqnarray}
satisfies Eq.~(\ref{maineqn}) where $d$ is as given in
Eq.~(\ref{ddef}) and $b$ and $a$ are two other dimensional,
integration constants. For an attractive force $b > 0$ and $d > 0$ but
$a$ can have any sign. If we assume that as $r\to \infty$, $V(r)\to 0$
then we get a relation between the constants $d$, $b$ and $a$ as
\begin{eqnarray} 
E=\sqrt{db^2+a}\,.
\label{dba}
\end{eqnarray}
From Eq.~(\ref{nvform}) we get the force acting on the particle,
$${\bf F}=-\nabla V(r)\,,$$
as
\begin{eqnarray}
{\bf F}=-\frac{d\left(b+\frac1r\right)}
{r^2\sqrt{d\left(b+\frac1r\right)^2+a}}\,\,\hat{r}\,,
\label{force}
\end{eqnarray}
From the form of the force and Eq.~(\ref{dba}) we immediately see that
if $a=0$ we have $b=E/\sqrt{d}$ and we get back the Newtonian or the
Coulombic potential. From the form of the potential as written in
Eq.~(\ref{plaw}) we can furthermore identify $\alpha=\sqrt{d}$ and 
consequently when $a=0$ we have $b=E/\alpha$.  If $a\ne 0$ then the form of the
force is non-trivial. The form of the force as given in
Eq.~(\ref{force}) cannot be reduced to the harmonic oscillator force
in any limits of the constants. This shows that special relativistic
effects do not allow stable circular orbits in the presence of a force
which is proportional to the negative of the displacement
vector. Although the force expression in Eq.~(\ref{force}) is
mathematically interesting but in physics we do not encounter such a
force, except the $a=0$ limit.

From the expression of $V-E$ as given in Eq.~(\ref{nvform}) we get
\begin{eqnarray}
J(u) &\equiv& \frac{(V-E)}{L^2c^2}\frac{dV}{du}
= \frac{d}{L^2c^2}(u+b)\,.
\label{jup}
\end{eqnarray}
yielding 
\begin{eqnarray}
\frac{d^2 u}{d\theta^2} +u= \frac{d}{L^2c^2}(u+b)\,,
\label{forbit}
\end{eqnarray}
which gives the orbit equation of the relativistic particle which is
acted on by a force given by Eq.~(\ref{force}). As in general
$d=L^2c^2(1-\zeta^2)$ for a stable closed orbit where
$\zeta$ must be a rational number, we get
\begin{eqnarray}
\frac{1}{r}=\frac{1}{R}\cos(\zeta \theta) + \frac{b(1-\zeta^2)}{\zeta^2}\,,
\label{prec}
\end{eqnarray}
where
\begin{eqnarray}
R=Lc\zeta\left[\frac{b^2L^2c^2(1-\zeta^2)}{\zeta^2}+a-m^2c^4\right]^{-1/2}\,.
\end{eqnarray}
The equation of the orbit in Eq.~(\ref{prec}) shows that in the most general case we will
have precession of the orbits dictated by the condition $(2\pi+\delta\theta)\zeta=2\pi$, which predicts that the 
orbit precesses by an angle
\begin{eqnarray}
\delta\theta=\frac{2\pi(1-\zeta)}{\zeta}\,,
\end{eqnarray}
per orbit. 
\subsection{The case of large perturbations}
Till now we have utilized first order perturbation from a circular
orbit as described in Eq.~(\ref{pertbj}) in the beginning of this
section. To include higher order perturbations from a circular orbit
we require more terms in the Taylor series expansion of $J(u)$ in
Eq.~(\ref{pertbj}). The second order effects will come from terms
proportional to $(d^2J/du^2)_{u_0}$. If the general form of the
potential $V(r)$ satisfies Eq.~(\ref{nvform}) then it is immediately
clear from Eq.~(\ref{jup}) that all derivatives of $J(u)$, except the
first, vanishes. Consequently in the relativistic case it is
impossible to restrict the constants $\zeta$, $a$ and $b$ by higher
order perturbation terms to the circular orbit. For higher order
perturbations from circular orbits the form the potential as given in
Eq.~(\ref{nvform}) remains the same.
\section{Connection of the present work with some related works}
\label{reltd}
One of the findings of the present article is related to the absence
of stable, circular orbits for relativistic particles in presence of a
harmonic oscillator potential. The trajectories of relativistic
particles in a three dimensional harmonic oscillator potential has
been studied previously by L.~Homorodean in
Ref.~\cite{homorodean}. The method followed in the referred work is
completely equivalent to the one followed in the present work. It is
interesting to note that in Homorodean's analysis the general shape of
the orbit in the relativistic case is not an ellipse, or a circle, but
a rosette shaped curve. In presence of the oscillator potential the
angular momentum of the orbiting particle with a specific energy has
an upper bound. The trajectory of the relativistic particle can only
be a circle when it has the highest angular momentum for a fixed
energy.  In Ref.~\cite{homorodean} the author does not give any
information about the stability of the orbits. In the non-relativistic
limit the orbit of the particle can be circular. In the light of the
findings in Ref.~\cite{homorodean} of Homorodean the prediction of the
absence of a stable, circular orbit in the oscillator potential is a
sensible result.

Bertrand's theorem in non-relativistic classical mechanics has
inspired some authors to propose a space-time (a metric to be precise)
where any bounded trajectory of a particle is periodic in nature. This
kind of a space-time is named as Bertrand space-time.  The works of
Perlick, Ballesteros, Enciso, Herranz and Ragnisco, in
Refs.~\cite{perlick, enciso}, try to generalize the results of the
classical Bertrand's theorem on a flat 3-space to a curved
3-manifold. In Ref.~\cite{perlick} the author found that a specific
form of a space-time which is asymptotically flat can support
Keplerian orbits. The asymptotically flat Bertrand space cannot support
closed trajectories expected in an oscillator type of potential. One
of the findings of the present article predicts that even in flat
space relativistic effects forbid closed, stable trajectories of
particles in presence of an oscillator potential.
\section{Conclusion}
\label{concl}
The outline of the article is based on the well known
Bertrand's theorem on central potentials and orbits of particles as
described in most of the classical mechanics books.  Like the
non-relativistic case the relativistic particle's orbit around a
potential source takes place in a plane where the angular momentum and
presumably the energy of the orbiting particle remains constant.  The
main difference between the non-relativistic orbits and relativistic
orbits crops up in the orbit equation itself. Unlike the
non-relativistic case in the relativistic case the orbit equation
depends upon the total energy of the particle.  The main aim of the
article was to find out possible forms of central potentials which can
produce stable circular orbits for relativistic particles. The
stability condition for the orbits can be transformed to a non-linear
differential equation for the central potential. It is seen that one
of the solutions of the non-linear differential equation for the
central potential is just the normal Coulomb potential. But relativity
affects the properties of the orbits by curtailing the angular
momentum values beyond a certain limit. Except the Coulomb potential
solution we find that the stability equation has another general
mathematically interesting solution which is unlike any potential
which we use in conventional physics.  In a specific limit the general
solution reproduces the Newtonian or Coulomb form. In the relativistic
version of the central force problem we lack some restrictions on the
potential which can produce stable circular orbits.  In the
non-relativistic version minimizing the effective potential one can
figure out the radius of the circular orbits and higher order
perturbation corrections to the stability condition of the orbits 
could be used for unravelling the exact nature of the potential. In
the relativistic version none of those restrictions remain and
consequently the general solution of the potential contains some
constants whose values cannot be analytically calculated. 

An important fact which comes out from the article is
about the non-existence of the harmonic oscillator potential as a
solution of the stability equation. In non-relativistic treatment of
the Bertrand's theorem it is well known that only two kinds of potentials
can produce stable circular orbits, one is of the Coulomb type and the
other is of the harmonic oscillator type. The Coulomb form of the
potential passes the stability test for circular orbits but the
harmonic type does not.

The article was focussed on some mathematical properties of
relativistic particle orbits in a central potential. Before we finally
conclude it is pertinent to say some thing on the practical side of
the relativistic central force problem. Atomic physics always remains
a store house of exciting phenomena and one of the places where one
may like to apply the tools of relativistic central force problems
lies inside the atom. This fact was discussed in Ref.~\cite{boyer}.
People have studied about the Schr\"{o}dinger equation and the Dirac
equation in presence of the Coulomb potential. It can be quite
interesting to study the analogous problems using the potential
presented in this article instead of the Coulomb potential. This
attempt can seriously shed some light on the physics of the atoms.

More over as there exists some work on the general relativistic
generalization of Bertrand's theorem one may expect that in the
simplest of the situations, where space-time remains flat, the results
of the present work can be applied for the orbits of very fast moving
bodies interacting via Newtonian gravity with a massive source.
\vskip .1cm
\noindent
{\bf Acknowledgement:}\,\,The authors acknowledge the illuminating
comments from H.~C.~Verma after he read the initial manuscript. Many
of his suggestions were implemented while preparing the final version
of the manuscript.


\begin{thebibliography}{999}
\bibitem{goldstein} H.~Goldstein, ``Classical Mechanics'', $2^{\rm nd}$ edition,
Narosa Publishing House, (1993)

\bibitem{boyer} T.~H.~Boyer, Am.~J.~Phys. {\bf 72}, 992-997, (2004)

\bibitem{tork} U.~Torkelsson, Eur.~J.~Phys {\bf 19}, 459-464, (1998)

\bibitem{reut} Z.~Reut, [Q.~J]~Mech.~Appl.~Math., {\bf 39}, 417-423, (1986)

\bibitem{frommert} H.~Frommert, Int.~J.~Theor.~Phys., {\bf 35}, 2631-2643, 
(1996)

\bibitem{landau} L.~D.~Landau, E.~M.~Lifshitz, ``The classical theory of
fields'', $4^{\rm th}$ edition, Elsevier, (2008)

\bibitem{jackson} J.~D.~Jackson, ``Classical Electrodynamics'', $2^{\rm nd}$
Edition, Wiley Eastern Limited, (1975). 

\bibitem{homorodean} L.~Homorodean, Europhys.~Lett., {\bf 66}, 8-13, (2004)

\bibitem{perlick} V.~Perlick, Class.~Quantum~Grav., {\bf 9}, 1009-1021, (1992)

\bibitem{enciso} A.~Ballesteros, A.~Enciso, F.~J.~Herranz and O.~Ragnisco,
Class.~Quantum~Grav., {\bf 9}, 165005 (13pp) (2008)

\end{thebibliography}
\end{document}